**Odors in olfactory bulb are defined by a short discrete temporal sequence: recognition by brute-force conversion to a spatial pattern (chunking)**

H. SANDERS[1], B. KOLTERMAN[2], R. SHUSTERMAN[3,5], D. RINBERG[3,4], A. KOULAKOV[2], J. LISMAN[1]

[1]Brandeis Universisty, Waltham, MA; [2]Cold Spring Harbor Laboratory, Cold Spring Harbor, NY; [3]HHMI Janelia Farm, Ashburn, VA; [4]New York University; [5]University of Haifa, Haifa, Israel.

**ABSTRACT:**

Mitral cells, the principal neurons in the olfactory bulb, respond to odorants by firing bursts of action potentials called sharp events. A given cell produces a sharp event at a fixed phase during the sniff cycle in response to a given odor; different cells have different phases. The olfactory bulb response to an odor is thus a sequence of sharp events. Here, we show that sharp event onset is biased toward certain phases of the ongoing gamma frequency oscillation. Thus, the signature of an odor is a discrete sequence. The fact that this sequence is relatively short suggests a new class of "brute force" solutions to the problem of odor recognition: cortex may contain a small number of modules, each forming a persistent snapshot of what occurs in a certain gamma cycle. Towards the end of the sniff, the collection of these snapshots forms a spatial pattern that could be recognized by standard attractor-based network mechanisms. We demonstrate the feasibility of this solution with simulations of simple network architectures having modules that represent gamma-cycle specific information. Thus "brute force" solutions for converting a discrete temporal sequence into a spatial pattern (chunking) are neurally plausible.

**INTRODUCTION:**

Recent work in awake rodents showed that odors are represented by a temporal code in the olfactory bulb (OB). Mitral cells, the output neurons of the bulb, generate "sharp events," which are high-frequency bursts of action potentials (up to 200 Hz) having a duration as short as 20 ms. The sharp events evoked by a given odor in a given cell occur at a precise phase with respect to the sniff cycle [1,2]. This phase is different for different cells, tiling the several hundred milliseconds of the sniff cycle. Thus, an odor is defined by a temporal sequence of sharp events.

An important unresolved issue is whether this sequence is continuous or discrete. Gamma frequency (30-100 Hz) oscillations are observed in the field potential of the olfactory bulb [3]. Such oscillations discretize population firing in other systems, such as the hippocampus (reviewed in [4]) and can affect spiking in mammalian olfactory bulb [5]. Studies in the insect antennal lobe have shown that odors are represented by a sequence of cell firing and that this sequence is discretized by fast oscillations [6]. It was therefore of interest to test whether the occurrence of sharp events is discretized by gamma oscillations. Our results show that this is the case. Thus, the signature of an odor in the olfactory bulb output is a discrete sequence.

We have gone on to consider the implications of this finding for the unsolved problem of how odor recognition occurs in olfactory cortex. Extensive theoretical work has gone into characterizing the process of recognition in attractor networks, which can classify spatial patterns of input [7–9]. If odors are to be recognized by an attractor network, the odor-specific temporal sequence of sharp events must be converted into a spatial pattern. However, there is little theoretical understanding of how a temporal-to-spatial conversion might be done. Several classes of solutions, such as the Reichardt detector [10], the tempotron [11], or the time delay neural network [12], require that the dynamics of the individual units (e.g., axonal conduction delays and membrane time constant) be on the same order as the duration of the sequence. However, these classes of solutions do not appear likely in olfaction because the duration of sequences is >100 ms, much longer than the dynamics of single neurons, which is on the order of 10 ms. The solution is therefore likely to depend on biophysical processes having a longer timescale. One class of such biophysical processes is those that underlie working memory, a form of memory that can maintain information for many seconds [13]. Here, we suggest how similar maintenance processes could be utilized by cortex to convert the representation of an odor from a temporal sequence into a spatial pattern. Specifically, we propose a brute force solution in which cortex contains a small number of modules, each of which has persistent activity [14,15] representing input that occurred during a specific gamma cycle. These snapshots, collectively, would provide a spatial pattern that captures the entire temporal pattern during a sniff. To make a module uniquely represent the nth gamma cycle, there must be a mechanism that prevents persistent firing before the nth cycle. Furthermore, inputs that arrive after the nth gamma cycle must be unable to produce persistent activity. It was therefore important to determine whether there are neurally plausible mechanisms for solving these problems. A solution might be of importance to the more general problem of sequence recognition (chunking) in other neural systems [16].

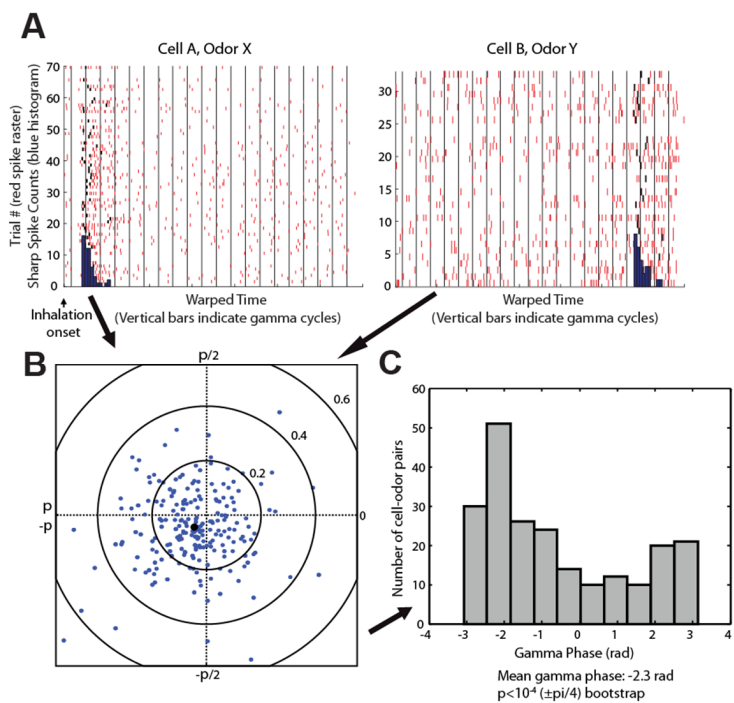

**Figure 1. Sharp event onsets are modulated by gamma oscillations, indicating a gamma-discretized sequence of sharp events.** **A)** Mutliple trials showing responses to an odor. Responses are shown to two cells (different odors). Red ticks are the spikes. Black ticks are spikes that were determined to be the first spike in a sharp event (see Materials and Methods). Each gamma cycle of each trial was warped so that the borders of gamma cycle are aligned with the vertical black lines. Onset histograms in warped time (blue). **B)** The average gamma phase of sharp event onsets is plotted in polar coordinates for each of 218 cell-odor pairs. The angle of each dot represents the preferred gamma phase of sharp event onset, and the distance of the dot from the origin represents the degree of synchronization over trials. The black dot is the average over all cell-odor pairs. **C)** Plot of phase preference for all 218 cell-odor pairs has a unimodal distribution that shows a statistically significant peaked distribution $p<10^{-4}$ (see Methods). **D)** Schematic: sharp events (ovals)have an onset biased toward a certain phase of a gamma cycle. One can threrefore think of the OB activity as a discrete sequence in which each item in the sequence is the ensemble of mitral cells that have sharp events during a given gamma cycle. Different odorants evoke different sequences of sharp events.

**RESULTS:**

**Onsets of sharp transient events are synchronized with the gamma cycle.**

Single unit recordings were made from olfactory bulb in the awake state to identify sharp events. The field potential was simultaneously recorded to measure gamma oscillations. Consistent with previous work [3,17], the power spectrum of the field potential showed a peak in the gamma frequency range around 60 Hz. We examined the synchronization of the onset of sharp events (time of first spike in a sharp event) with the simultaneously occurring gamma oscillations. Examining 218 cell-odor pairs (two examples shown in Fig.1A), we found that sharp event onset was modulated by gamma phase. The gamma phase of sharp event onset over the entire population was significantly biased toward a certain phase of the gamma cycle (mean gamma phase -2.3 rad, $p < 10^{-4}$ ($\pm \mathrm{pi}/4$) bootstrap, Fig.1C). We found no dependence of preferred gamma phase on the time of the sharp event during the sniff cycle (not shown). Overall, our findings indicate that the onsets of sharp events are biased toward a certain phase of the gamma frequency.

This finding leads to a description of athe output of the olfactory bulb not as a continuous temporal sequence, but rather as a discrete temporal sequence organized by gamma frequency network oscillations (Fig.1D). The number of such gamma cycles relevant for recognition may depend on the complexity of the recognition task, but in any case, it cannot be large, given that recognition can occur in less than a sniff cycle [18]. For gamma cycles of about 15-20 ms, and given the fact that odor identification can often occur in less than 100-150 ms of neural processing time [19,20], less than 10 gamma cycles define an odor sequence.

**Algorithm for sequence recognition**

How might an odor-specific discrete sequence be recognized as a unitary odor? The best-studied basis of recognition is the classification of spatial patterns by attractor networks [7,8]. In fact, attractors have been suggested to be involved in olfactory identification [21]. These networks select the stored memory closest to the input pattern and can do so even in the presence of noise or incomplete patterns. However, for odor recognition to be based on such a mechanism, the temporal sequence that emerges from the olfactory bulb (OB) must be converted into a unique spatial pattern to be provided to the attractor network. The discretization of odor-specific sequences by a relatively small number of gamma cycles allows for the possibility of a brute-force mechanism for accomplishing this: that the olfactory cortex contains a corresponding number of discrete modules, each specialized to produce a persistent snapshot of what occurred in a specific gamma cycle ("gamma cycle specificity"). Because these representations are persistent, a spatial pattern will evolve during a sniff, as each successive gamma cycle comes to be represented by the activity in successive modules.

Here, we explore two related implementations of this algorithm. In both cases, cortex is composed of several modules, each of which receives all OB input at all times (Fig.2A and Fig.3A). Both models also employ bistable units to produce persistent activity after appropriate activation. The first model is a network built of binary neurons, in which gamma-cycle specificity of each module is due to differences in activation thresholds across modules. The second model is a more biophysically detailed model consisting of networks of spiking neurons. In this model, gamma-cycle specificity results from bistability in each module being conditional on receiving feed-forward input from the immediately preceding module.

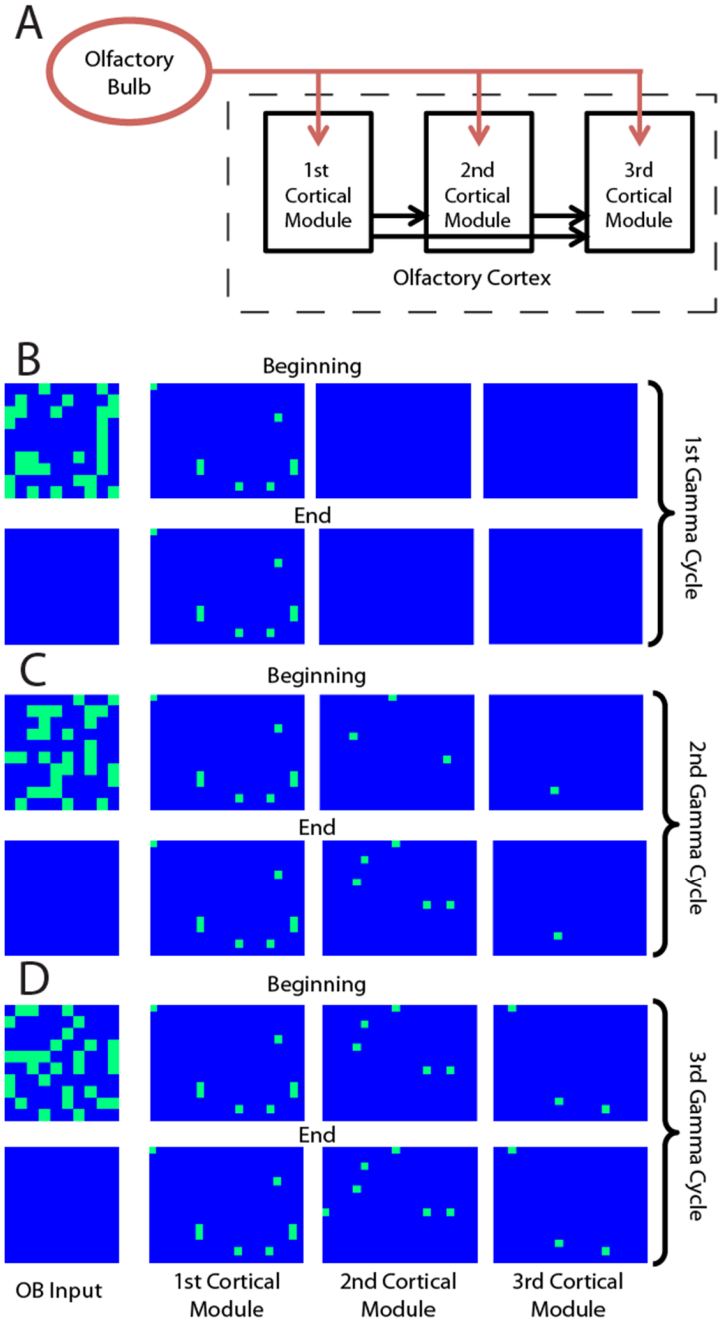

**Figure 2. The binary neuron model for sequence representation. A)** Schematic of the model. Three distinct modules in PC are tuned so that they show persistent activity starting in different gamma cycles. All units receive a similar number of random connections from OB (red arrows). The modules have different thresholds, and the modules with lower thresholds, once activated, provide feed-forward input to all modules with higher thresholds (black arrows). The first module (which has the lowest threshold) will, after activation, prime the second module, and so on. The modules also contain random recurrent connectivity among their units (not shown, see Methods). **B-D)** Each panel shows the activity in the OB (left) and olfactory cortex at the beginning and the end of a given gamma cycle. Each rectangle on each panel represents the activity of a single unit, where blue represents non-firing and green represents firing. During each gamma cycle, some subset of mitral cells are activated as a consequence of sharp events during that cycle and fire persistently. Because of the gradient of excitability, later modules only begin to fire during later gamma cycles in the sniff cycle.

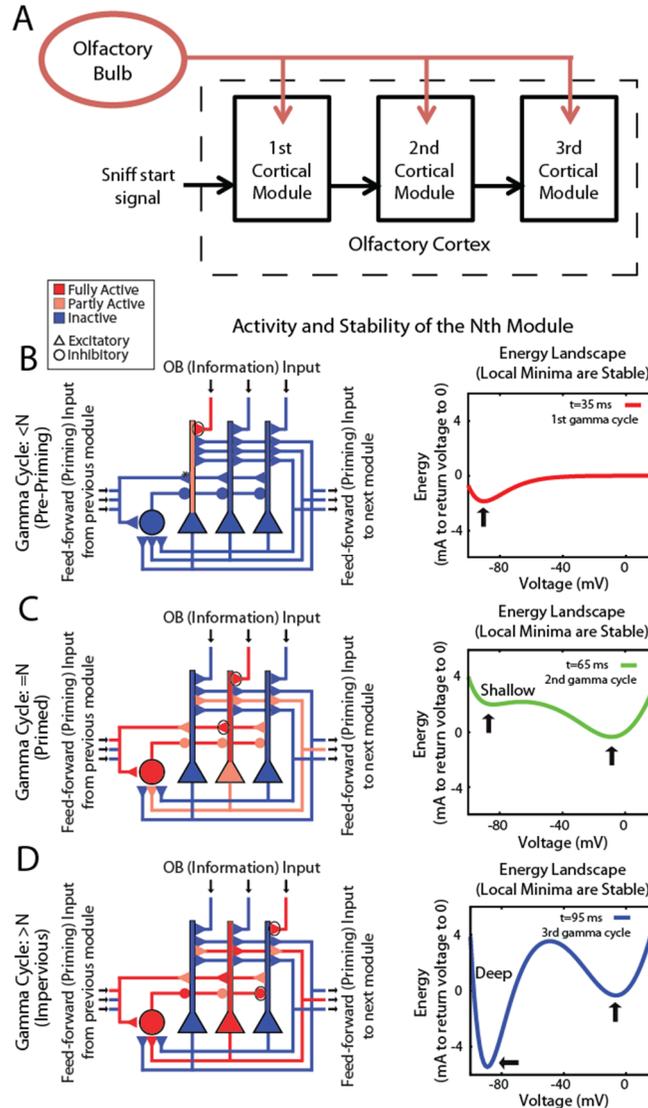

**Figure 3. Spiking network model architecture for sequence generation. A)** OB input is provided to all modules. Each module receives feed-forward excitatory input from the previous module onto its excitatory and its inhibitory cells, which together prime bistability in the excitatory cells. **B-D)** On the left, a schematic of the nth module shows its behavior on a given gamma cycle. Three excitatory (triangle) and one inhibitory (circle) cells are shown. Blue represents hyperpolarization or lack of activity, whereas red represents depolarization or spiking. Pink represents moderate depolarization or synapses that are active but not passing much current because of their voltage dependence. On the right, an energy landscape of neuron 200 in the 2nd module in the simulation shown in Fig.5 in order to demonstrate the voltage stability of a cell before, during, and after its OB input is active during the correct gamma cycle. Local minima represent stable voltages. **B)** Before the nth gamma cycle, the nth module has not received feed-forward priming and thus is not bistable, as can be seen by the fact that there is only a single energy well in the right panel (this state is determined by leak conductance and basal KIR conductance). Any OB input during such a gamma cycle can produce firing, but this will not be maintained. **C)** The nth gamma cycle is the first gamma cycle during which the nth module receives feed-forward input activates the NMDA conductance and slowly developing KIR (GABA-B) conductance, priming its bistability (note two stable voltages in the right panel). Because the hyperpolarized energy well is shallow, cells that receive OB input during that gamma cycle can switch to the depolarized stable state. **D)** After the nth gamma cycle, slow GABA-B-activated voltage-dependent inhibition has built up enough to make the hyperpolarized state very stable (as shown by the deep energy well), preventing cells that receive subsequent OB input from switching to the depolarized stable state. Therefore, their activation is not persistent.

**Binary neuron model**

The architecture and behavior of this model are shown in Fig.2. It is assumed that olfactory cortex is divided into a small number of ordered modules, each of which contains many neural units. In addition to input from many mitral cells, units in each module receive excitation from all previous modules (Fig.2A) and feedback inhibition from units in their own module (not shown). Units in each module are bistable; meaning that for some range of inputs they can be either active or inactive depending on its history. There is a threshold level of input that causes the transition from the inactive to the active state. Once this transition occurs, the unit can remain active even when the input falls below the threshold level. This implementation is a simplification of the processes likely to produce the persistent firing in various brain regions [22,23]. Among the modules, there is a gradient of threshold, so that later modules need more input to make the transition to the active state. Part of input needed to reach threshold can come from the active cells in earlier modules. For example, the units in the second module are initially not excited enough by OB input alone to become persistently active. However, after activation of the first module and the resulting feed-forward excitation (priming), units in the second module can be persistently excited by the OB input that occurs during the second gamma cycle. The gradient of threshold enforces the requirement that OB activity before the nth gamma cycle not affect the nth module. The requirement that OB activity after the nth gamma cycle not affect the nth module is enforced by the intra-module feedback inhibition; after the nth gamma cycle, there is so much local inhibition that OB input is not sufficient to trigger further transitions from the down state to the up state. Cells already in the upstate can, however, continue to fire. This persistent firing thus represents a snapshot of the consequences of OB during the nth gamma cycle.

We term this the binary model, because units are either active or not. This model is an adaptation of a previously published network model of integration [24]. Inspection of the network activity (Fig.2B-D) shows that what happened during the first gamma cycle is retained by the first module until the end of the sniff. Similarly, what happened during the second gamma cycle is retained by the second module. This and corresponding actions in the other modules ensure that the final spatial representation depends uniquely on the temporal sequence of gamma-cycle-specific mitral cell outputs.

**Spiking neuron model**

The above model includes several simplifying assumptions including that units are either on or off; moreover, bistability is taken as a given without including a specific conductance mechanism. It was therefore of interest to develop a model in which bistability arose from specific mechanisms. To explore this possibility, we constructed a series of modules, each of which was composed of Hodgkin-Huxley conductance-based spiking neurons. We incorporated a previously postulated form a robust bistability arising from the interaction of NMDAR and GABA-B-activated KIR [25]. In the binary model, gamma cycle specificity is due to differing activity thresholds in the different modules. In contrast, in the spiking model, each module is biophysically identical, but sequential activation is conferred by asymmetric, feed-forward excitation necessary for the existence of bistability.

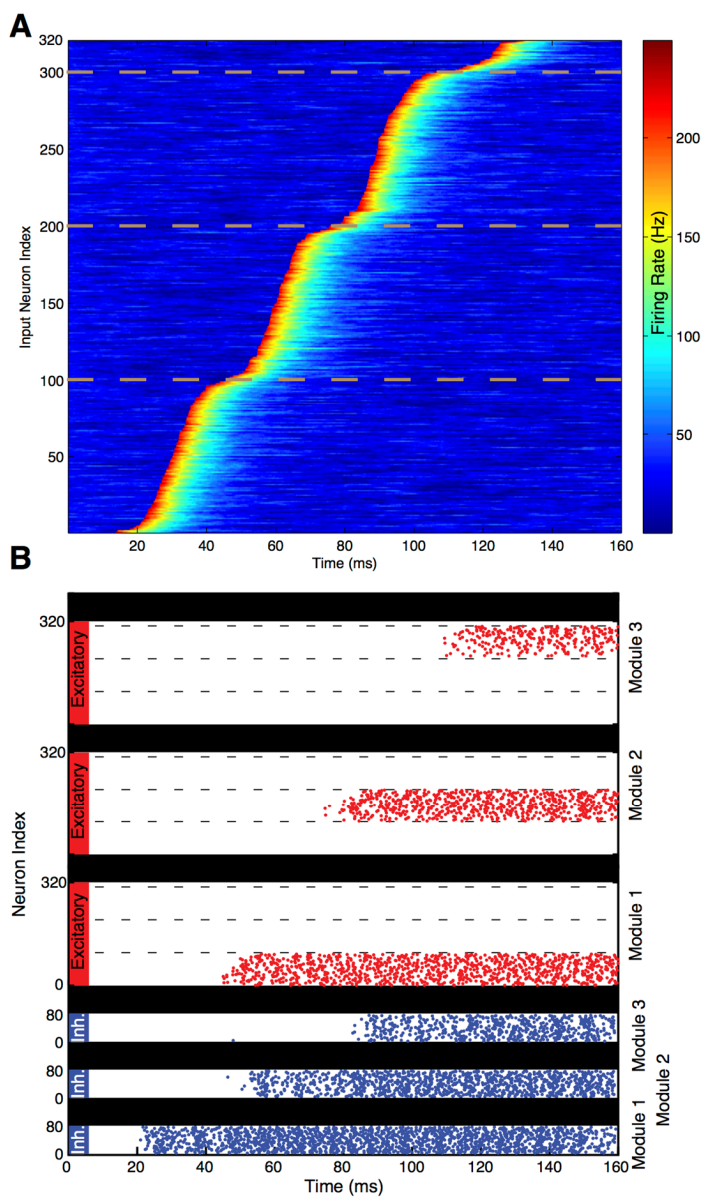

**Figure 4. Illustration of spiking network performance. A**) Firing rate of the 320 OB cells during the simulation. Different sharp events begin in different gamma cycles. The dotted lines demarcate the three groups of 100 cells that have sharp events in each of the three gamma cycles shown. OB cells are ordered by sharp event latency for ease of visual inspection. **B**) Each excitatory cell (red) receives input from the single OB cell with the same index in this simplified simulation. Each module successfully maintains a record of the OB cells that provided input during a particular gamma cycle. As shown, the 80 cells activated during each gamma cycle are sequentially numbered, but the pattern of activated cells could be arbitrary. Spikes of the 80 inhibitory interneurons in each module are shown in blue. Interneurons start spiking before excitatory neurons because of the feed-forward excitation that they receive. During the last 20 ms, the network activity has a spatial pattern that reflects the pattern of mitral cell activity during the first three gamma cycles.

Each module goes through three stages during the course of the sniff. Before the gamma cycle to which that module is dedicated (the nth gamma cycle for the nth module), there is only one stable voltage for the neurons in that module. That voltage is determined primarily by leak and constitutively active KIR. Therefore, the voltage of any neuron receiving OB input decays to rest soon after that input is removed (Fig.3B). Towards the end of the n-1 cycle there will be feed-forward input from the n-1 module that increases the AMPA and NMDA conductances on all excitatory and inhibitory cells in the nth module. The activation of the NMDA conductance in the feed-forward projection "primes" the nth module by creating the potential for a stable depolarized voltage in the dendrites of excitatory cells [26]. Thus, cells receiving OB input during the nth gamma cycle are able to switch into a stable depolarized state (Fig.3C). The feed-forward projection also excites the inhibitory cells in the nth module, which leads to slower increase of GABA-B-activated KIR conductance during the next 20-80 ms [27] in the excitatory cells in the module. This eventually makes the hyperpolarized state very strongly stable (note deep energy well in Fig.3D). The strength of this stability prevents cells in the hyperpolarized state from activating due to OB input that occurs in the n+1 and later gamma cycles, locking the activity state of the module and making it impervious to further input (Fig.3D). Thus, the network operates in such a way that persistent firing pattern in a given module is determined by OB input only during a given gamma cycle. Fig.4 shows that a network based on these principles indeed maintains a spatial pattern that is observable at the end of the sniff and that each module represents the input that occurred during a specific gamma cycle.

Does this sequence-decoding mechanism depend crucially on the fact that OB input is discrete? To examine this question, we varied the extent of discretization of the temporal sequence and determined how this affected the accuracy of the spatial pattern produced in our cortical network, quantified using an "accuracy index". This is based on a comparison of the activity of the network in the last 50 ms of the simulation to the activity that would be expected if activity of a OB cell in a particular gamma cycle led to activity of the cortical cell of the same index in the module dedicated to that gamma cycle (a positive accuracy index means that the network gets more excitatory cell activities right than it got wrong). In each gamma cycle, the variability of sharp event onset in the population is described by the width of a normal distribution, so we varied the extent of discretization by changing the width of the distribution of sharp event onsets. We ran simulations over a range of several parameters (NMDA, GABA-A, and GABA-B maximal conductance and number of OB cells activated each gamma cycle) and plotted in Fig.5 simulations having a parameter set that was successful for at least one gamma discretization level (accuracy index>0.7). Fig.5 shows that, as the width of the normal distribution was increased, there was reduced ability of the simulated cortical network to form an accurate spatial representation of the sequence. Thus, the discretization of temporal input is important for effective decoding of sequences by a modular receiver.

## DISCUSSION

The rodent olfactory bulb output has been shown to use a temporal code to represent odors [1]. In fact, rats are able to perform a task that depends on the specific timing of channel-rhodopsin-induced activity relative to the sniff cycle [28], and individual neurons in piriform cortex can be sensitive to the relative timing of OB neuron activations [29]. Here we show that this temporal code is not continuous but is rather formed by a sequence in which sharp event onset occurs at a preferential phase of the ongoing gamma oscillation. Evidence for a similar discrete gamma code has been presented for locust olfaction [30]. Thus, odor representation by a discrete temporal sequence may be a general property of olfactory recognition.

We have presented two related models for how discrete temporal sequences might be converted to spatial representations. In these models, there are separate cortical modules; each persistently represents information relating to olfactory bulb output during a specific gamma cycle. This gamma-cycle specificity has two requirements. One is that OB activity *before* the designated cycle for that module (the nth cycle) does not produce persistent activity. This is solved in the binary neuron model with a gradient of threshold (for transition to the up state) in different modules requiring that all previous modules be activated before the next module can reach the activation threshold (Fig.2). In the spiking neuron model, the bistability is not intrinsic but requires feed-forward "priming" of bistability by activity in the previous module before incoming OB input can be retained (Fig.3-4). The other requirement for gamma-cycle specificity is that OB activity during gamma cycles *after* its designated cycle not affect a module. This was achieved in both models because inhibition became strongly elevated after the nth cycle (Figs. 2-4), preventing OB from triggering further transitions to the up-state. Our main conclusion is that networks in which gamma cycle specificity of the individual modules are biologically plausible. These modules represent a collection of gamma-cycle specific snapshots of OB activity, providing a complete picture of the input sequence. We term these brute-force solutions because a separate group of cells is dedicated to each gamma cycle.

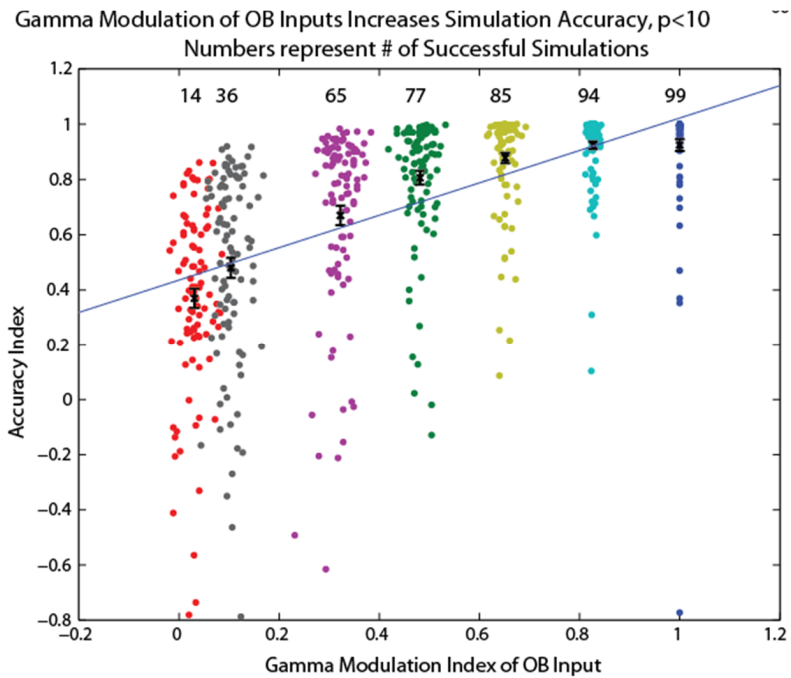

**Figure 5. Gamma modulation of OB inputs increases spiking network accuracy.** The network simulation was run over a range of parameters (see Methods) for different sharpnesses of gamma modulation of OB input. The sharpness of gamma modulation was defined by the width of the gaussian distribution from which the timing of onset of the sharp events was drawn for each gamma cycle. Each dot in the figure represents a simulation and is associated with the sharpness of gamma modulation (color), the measured gamma modulation index (x-axis), and the accuracy of the simulation (y-axis). The accuracy index quantified how similar the final network state was to the state that would be expected if each module perfectly maintained the OB activity of its gamma cycle (see Methods). If a parameter set had a successful simulation (accuracy index>0.7) for any extent of gamma modulation, the simulations with that parameter set are shown here for all sharpnesses of gamma modulation (112 total parameter sets shown). The number of successful simulations and the average accuracy index both significantly increased with increasing gamma modulation of the OB inputs.

An important issue is the layout of cells in different modules. At one extreme is a non-topographic model in which modules are defined completely by intracortical wiring among random cells. At the other extreme is topographic mapping in which all of the cells within a local region would contribute to the same module and different modules would be laid out in order along an axis of cortex. At this point, neither possibility can be excluded. We entertain the possibility that the conversion of the temporal code to a spatial code might occur in the first-order olfactory cortex (piriform cortex). It is noteworthy that asymmetric, feed-forward connectivity of the type assumed in our second model is evident among subdivisions of olfactory cortex [31], suggestive of a functional gradient across this axis. The observed random projection from OB to PC [32] is *not* inconsistent with a functional gradient because all of the modules in our model receive similar inputs from OB. However, a critical prediction of our models is that cells demonstrate odor-evoked persistent firing. Recordings from primary olfactory cortex show many cases of transient firing but also a few examples of persistent firing (Fig. 2E of Miura et al., 2012). We emphasize that cases of transient firing are not unexpected; this could occur when input transiently excites a module whose bistability is not yet primed by the previous module. Nevertheless, the paucity of persistent firing in piriform raises questions about whether this is the site where temporal to spatial conversion occurs. Another possibile site is the endopiriform nucleus, which is deep to piriform cortex and is sometimes called layer 4 of piriform cortex [31]. Importantly, neurons in the endopiriform show long-lasting (>100ms) depolarizations in response to afferent stimulation [34] similar to those required for our model. Furthermore, much research on the endopiriform because of its involvement in epilepsy [35], where persistent firing becomes pathological. It would thus be very interesting to determine whether odors can produce persistent firing in the endopiriform nucleus.

Our model makes a very specific prediction about the expected properties of persistent firing: for a given cell, the onset of persistent firing will occur at a specific phase of the sniff cycle irrespective of the odor identity. If the modules are topographically organized, this would create a cortical wave which might be detectable in the field potential (for other examples of cortical waves, see [36]).

There have been several previous models for how a network can identify a temporally extended input sequence. One class of models depends on delays such that the beginning and end of a sequence reach a classifier at the same time [10,12,37–39]. Our model shares the idea that the beginning of the sequence is provided to the classifier at the same time as the end of the sequence. However, in our model, the input is available to the classifier throughout the sequence so that classification can begin to occur as soon as evidence accumulates instead of having to wait until the end. Additionally, other sequence decoding models do not employ biophysical mechanisms to generate delays of the length necessary to identify olfactory sequences with durations on the order of 100 ms. Still another class of models for sequence recognition is that of liquid/echo-state networks [40,41]. These networks rely on recurrent excitation to produce a trace that lasts for longer than the input but that nevertheless decays over time. Though such a system is useful for continuous inputs, the traces produced are necessarily approximate because of decay. In contrast, the type of networks that we have analyzed can convert temporal patterns to spatial patterns that are not subject to rapid decay. Such networks, though not appropriate for continuous inputs, seem ideally suited for the type of short discrete sequences that represent an odor.

The importance of gamma oscillations in sequence representation has been explored. One function of gamma is the synchronization of inputs, which allows for linear or non-linear coincidence detection processes [16,42]. On the other hand, gamma discretization allows for some level of noise in the precise timing of input activity as long as spikes stay within the same gamma cycle [43]. The functions of gamma discretization have been explored in the olfactory system itself and have been shown to be of benefit in disambiguation of similar stimuli [44]. Our results further support this conclusion. As shown in Fig.5, increased gamma modulation of the input directly correlates with the accuracy of representation in our network. We suggest that the function of gamma demonstrated here is to take a sequence that is too long for easy continuous analysis and to divide it into discrete elements that can be effectively dealt with individually.

There are other brain processes in which temporal sequences must be processed, including hippocampal replay [45,46] and language processing [47]. In psychology, the general problem of recognizing a temporal sequence and ascribing to it a single label is called chunking. Our model coverts a temporal sequence to a stable spatial pattern and thus is a potential mechanism that could underlie chunking.

## MATERIALS AND METHODS

The experimental procedure is as described in [1].

### Sharp spiking events

Mitral cell spike trains were ordered by session and split into unit-odor pairs. The spike trains for each trial were analyzed from a four-second window centered around the inhalation onset immediately following odor presentation. The PSTH was calculated by placing the spike trains for a given unit-odor pair in 10 ms bins and averaging over trials. Unit-odor pairs were then selected for further analysis if their trial-averaged responses contained "sharp spiking events." Sharp spiking events were defined as an increase in the PSTH with a peak at least 4.5 standard deviations ($\sigma$) above the baseline rate determined over the 2 s interval before odor presentation. The "sharp spike onset" is defined similarly to [1] as the time of the spike preceding the first inter-spike interval below a threshold of 1.5/maxFR (the maximum value of the PSTH) within a window around the peak of -2/maxFR to +4/maxFR. The timing of this spike was saved for each trial in the unit-odor pair's session.

**Gamma Phase Determination**

The LFP recorded from the same electrode as the unit being analyzed was filtered in Matlab with a zero-phase, 4-pole Butterworth bandpass filter between 40 and 80 Hz to isolate the gamma oscillation. The phase of a sharp spike onset ($\theta$) for a given trial was estimated using the LFP signal ($S$) and the timing of the first spike in the sharp event in a given trial ($t_{sp}$) described above. The temporal length of $S$ was restricted to the window around the peak of the PSTH as defined above. The phase was then calculated using $\theta(t_{sp}) = \arctan\left[\frac{\frac{dS}{dt}|_{t=t_{sp}}/\sigma_{dS/dt}}{S(t_{sp})/\sigma_S}\right]$.

Here the time derivative of the LFP, $\frac{dS}{dt}$, was calculated using the discrete gradient function in Matlab. The mean phase angle over trials was determined using the circular mean of phases as in [48]. For each unit-odor pair, a complex number $Z = N^{-1}\sum_{j=1}^{N}e^{2i\theta_j}$, which leads to the following results: $|Z|$ is the degree of synchronization over trials with 1 being the maximum, and $\arg(Z)$ is the gamma phase of the sharp event onset averaged over trials. Complex variable $Z$ for several cell-odor pairs is shown in Fig.1B, and the argument of $Z$ is shown in Fig.1C as representing sharp event onsets.

The mean phase across unit-odor pairs can then be found by averaging the $Z$ values. To determine the p-value of the mean phase across unit-odor pairs, we used the random sampling with replacement method. A set of size equal to the original number of unit-odor pairs was created by randomly choosing $Z$ values of unit-odor pairs out of the original set, allowing for repeats. A new value of the mean angle was calculated from this distribution and saved. This procedure was repeated 10,000 times. The fraction of these bootstrapped mean phase angles falling outside of a window $\pm\pi/4$ of the mean of the original distribution gives an estimate of the p-value.

**Binary neuron model (Fig.2)**

To test generality of our model, we used a simple binary neuron model that included neurons with bistability in their response without specifying the origins of bistability. The model was based on random and sparse connectivity between the bulb and the cortex, as well as within the olfactory cortex as detailed below. Our simulation included 100 OB mitral cells and three cortical modules containing 300 neurons each. OB mitral cells connected randomly and sparsely to cells in all of the olfactory cortex modules with a 1% probability. Neurons in each cortical module formed random sparse excitatory and inhibitory recurrent associative connections to other cells within their own module with 1% and 30% probabilities, respectively. These neurons also formed random excitatory connections with the subsequent module with 1% probability. Within modules, non-zero excitatory/inhibitory connections had the following values of strengths: $W_{ij}^{ex} = 0.1$ and $W_{ij}^{in} = -1.5$, respectively. Non-zero projections from the olfactory bulb, i.e., mitral cells, were $W_{ij}^{ex} = 2$, while the connections between modules were $W_{ij}^{ex} = 4$.

The state of each cortical neuron was defined by the input that this neuron receives $u_i$ that satisfied the equation $\tau \frac{du_i}{dt} = \sum_{j=1}^{N} W_{ij}^{ex} f_j + \sum_{j=1}^{N} W_{ij}^{in} f_j - \Delta_i$. Here $\tau = 20$ is the time constant and $\Delta_i$ is the offset that determined excitability of this neuron. The offsets for three cortical modules were 0.5, 2.5, and 4.5. The activation state for each cortical neuron had a hysteretic dependence on its inputs $f_i = F_{\pm}(u_i)$. The activation function $F_{\pm}$ was single valued for values of input variable $u$ satisfying $u > u_+ = 1$ and $u < u_- = -10$. For these values of parameters, the activation function $F_{\pm}$ was equal to 1 and 0 respectively. Within the bistable range, i.e., for $u_- \le u \le u_+$, $F_{\pm}$ was bistable and remained constant depending on prior history. Therefore, if a neuron was activated, the activation function within the bistable range remained equal to 1 whereas for an inactivated neuron, the activation function was 0. The activation occurred when inputs exceeded $u_+$, and inactivation happened when inputs fell below $u_-$.

The simulation was carried out over three gamma cycles using Runge-Kutta method with the time step $\Delta t = 0.2$. Each gamma cycle was split into two parts lasting 8 time units each. During the first part, the mitral cells sent inputs to the olfactory cortex (sharp events). To produce these inputs, we generated random variables $f$ =0 or 1 for each mitral cell that determined its activation state. During each gamma cycle 30% of mitral cells emitted the sharp events. The identities of responding mitral cells did not overlap between different gamma cycles. During the second part of the gamma cycle, the mitral cells were silent, i.e., their activation states were zero.

**Spiking network architecture**

The spiking network model is taken from [25] with minor modifications.

The networks used in this study contained three modules of $N = 400$ neurons each, of which $N_p = 320$ were excitatory and $N_I = 80$ were inhibitory. There were 320 external OB channels characterized by a firing rate, each of which synapsed onto one excitatory cell and all inhibitory cells. All excitatory neurons synapsed onto all other excitatory neurons in the same module with weight $w_{EE}$ and onto all inhibitory neurons in the same module with weight $w_{EI}$ as well as all excitatory neurons of the next module with weight $w_{ffE} = w_{EE}$ and all inhibitory neurons of the next module with weight $w_{ffI} = w_{EI}$. All inhibitory neurons synapsed onto all excitatory neurons within their module with weight $w_{IE}$. See below for values. The first module received spiking input from 320 sources meant to represent a sniff start signal, to be described in detail later. This architecture is shown schematically in Fig.3.

**Spiking neuron model**

The model neurons used in this study were Hodgkin-Huxley-type conductance-based neurons, modified from [25], who modified from [26]. The excitatory neurons had two compartments: a dendrite with voltage $V_d$ and a soma with voltage $V_s$. Separating the spike-generating conductances from the bistable synaptic compartment allows bistability to be maintained during the large somatic voltage fluctuations associated with action potential generation.

The dynamics of the compartmental voltages and the leak, noise, AMPA, NMDA, GABA-A, and GABA-B conductances are the same as [25] except for the following changes: 1) the maximal conductances were changed, and the ones used in this study are shown in Table 1; 2) the implementation of the external (OB) input was changed, and is described below; 3) $\alpha$ in the equation for the NMDA dynamics was changed from 0.1 to 0.5 because NMDA conductance activation required too many spikes ($\alpha$ had an original value of 1 in [26] but was changed in [25] to reflect the lack of saturation of the NMDA conductance with single spikes [49]); 4) instead of 25% of the KIR conductance being constitutively active, 5% of the KIR conductance was constitutively active and 95% was activated by GABA-B, giving a modified equation describing the KIR current: $I_{GABA_B/KIR} = g_{GABA_B/KIR}(0.05 + 0.95\sum_i s_i)\frac{V_d - E_{GABA_B/KIR}}{1 + \exp(0.1(V_d - E_{GABA_B/KIR} + 10))}$ (see [25] for definition of all variables).

| Synaptic | maximal $g$ per synapse (mS/cm$^2$) | | Reversal |
|---|---|---|---|
| Conductance | onto p-cells | onto I-cells | Potential |
| AMPA | $1.125/N_p$ | $1.125/N_p$ | 0 mV |
| NMDA | $4.5/N_p$ | $0.3/N_p$ | 0 mV |
| OB input | 0.4 | $0.2/N_p$ | 0 mV |
| GABA$_A$ | $0.2/N_I$ | 0 | -70 mV |
| GABA$_B$ | $130/N_I$ | 0 | -90 mV |

Table 1: **Values of maximal synaptic conductances used in simulations.**

The maximal conductance values in this table apply to the simulation shown in Fig.4. The simulations in Fig.5 all had different values of the following maximal conductances: $gAMPA$ onto p-cells and onto I-cells, $gNMDA$ onto p-cells, $g_{GABAA}$ onto p-cells, and $g_{GABAB}$ onto p-cells. The maximal conductance values used in the simulations in Fig.5 are described below under "Spiking network simulations."

The external synaptic input to a given cell in the network $I_{syn,ext} = s_{input}g_{input}(V - E_{syn})$, where $g_{input}$ is the maximal conductance given in Table 1 and the synaptic activation $s_{input}$ is the fraction of that conductance that is activated. $s_{input} = 0.00175r$, where the firing rate $r$ of the OB input was governed by dynamics given by $\frac{dr}{dt} = \frac{P}{\Delta t}\delta(t - t_{sp}) - \frac{r - r_{base}}{\tau_{sharp}} + noise$, where the peak firing rate $P = 200$ Hz, $\delta(t - t_{sp})$ is the Dirac delta function centered on the time of the sharp event peak for that mitral cell, the baseline firing rate $r_{base} = 20$ Hz, the time constant of decay for the firing rate of sharp events $\tau_{sharp} = 10$ ms, and $noise$ is drawn independently for each time step from a normal distribution with a mean of 0 and a standard deviation of 5 Hz/$\sqrt{\Delta t}$. The factor of 0.00175 in the calculation of synaptic activation is taken from the steady-state synaptic activation: $s(r) = \frac{\tau r}{1000}\frac{\alpha \exp(\frac{1000}{\tau r}) - \alpha}{\exp(\frac{1000}{\tau r}) - (1 - \alpha)}$, where the time constant of decay of synaptic activation $\tau$ is taken to be 2 ms and the fraction of unactivated receptor activated with a single spike $\alpha$ is 0.9. This function is practically linear, with a slope of ~0.00175, for firing rates <150 Hz.

For each mitral cell input channel, a sharp event onset time $t_{sp} = \lfloor i/N_{prf}\rfloor t_{gamma} + noise$, where $i$ is the index of the mitral cell, $N_{prf}$ is the number of cells active in each gamma cycle, $\lfloor\ \rfloor$ is the floor function, and the length of a gamma cycle $t_{gamma} = 30$ ms, meaning that the first term gives the time at the middle of the gamma cycle it was supposed to be active on. $noise$ is drawn from a normal distribution with mean 0 and standard deviation of 10 ms for Fig.4 and standard deviations of [0,1.5,3,4.5,6,9,12] in Fig.5. The mitral cells were then sorted by $t_{sp}$ and re-indexed.

As for the "sniff start signal" received by the first module, there were 320 channels, each of which projected to all excitatory and inhibitory cells in the first module with maximal synaptic conductances equal to those of the feed-forward projections from module to module. Each of the 320 sources spiked at a random time taken from a uniform distribution between 15 and 30 ms after the beginning of the simulation and every 10 ms thereafter until the end of the simulation. The AMPA and NMDA synaptic activations onto all neurons in the first module from a given channel were simulated by $s_x(t + \Delta t) = s_x(t)\exp(-\Delta t/\tau_x)$, and if the channel spiked during that time step, then $s_x(t + \Delta t) = s_x(t) + \alpha_x(1 - s_x(t))$, where $x$ = AMPA or NMDA, $\tau = 2$ ms, $\tau_{\text{NMDA}} = 100$ ms, $\alpha_{\text{AMPA}} = 0.9$, and $\alpha_{\text{AMPA}} = 0.5$.

**Spiking network simulations**

For Fig.5, the simulation was run repeatedly for all combinations of the following five parameters: maximal $g_{NMDA,EE}$ for excitatory-excitatory connections (both within module and feed-forward from one module to the next) from the set $[50\ 54\ 58\ 62\ 66\ 70\ 74\ 78\ 82\ 86\ 90]/(20N_p)$ mS/cm$^2$, maximal $g_{\text{GABA}_A}$ for inhibitory-excitatory connections from the set $[0.5\ 1\ 1.5\ 2\ 2.5\ 3\ 3.5\ 4]/(10N_l)$ mS/cm$^2$, number of mitral cell channels active on each gamma cycle $N_{prf}$ from the set $[0.10\ 0.13\ 0.16\ 0.19\ 0.22\ 0.25]*N$, maximal $g_{\text{GABA}_B}$ for inhibitory-excitatory connections from the set $[600\ 700\ 800\ 900\ 1000\ 1100\ 1200\ 1300]/(10N_l)$ mS/cm$^2$, and the standard deviation of $noise$ in the calculation of $t_{sp}$ from the set $[0\ 1.5\ 3\ 4.5\ 6\ 9\ 12]$ ms. Maximal $g_{AMPA,EE}$ for excitatory-excitatory connections (both within module and feed-forward from one module to the next) was calculated from $g_{\text{NMDA},EE}$ by $g_{AMPA,EE} = g_{\text{NMDA},EE}/4$. Maximal $g_{AMPA,EI}$ for excitatory-inhibitory connections (both within module and feed-forward from one module to the next) was set as $g_{AMPA,EI} = g_{AMPA,EE} = g_{\text{NMDA},EE}/4$.

The gamma modulation index $GMI = 1 - \sqrt{12}\sqrt{\sum_{i=1}^{N}(\phi - 0.5)^2/N}$ is basically the standard deviation of sharp event onset phases from the cycle midpoint, normalized by $1/\sqrt{12}$ (the standard deviation from the cycle midpoint of uniformly distributed events) and subtracted from 1 in order to give an index that ranged from -1, representing all events occurring at the cycle edge, to +1, representing all events occurring at the middle of the cycle. A gamma modulation index of 0 would be arrived at with a uniform distribution of events.

The accuracy index $AI = 1$ − the failure index. The failure index is calculated based on the activity of the network in the last 50 ms of the simulation. The failure index is (the number of active excitatory cells not corresponding to sharp events during their gamma cycle + the number of silent excitatory cells corresponding to sharp events during their gamma cycle) divided by the number of sharp events that occurred during the first three gamma cycles. An accuracy index of 1 represented that excitatory cells in the network were active if and only if their mitral cell channel had a sharp event during their gamma cycle. An accuracy index>0 meant that the network got more excitatory cell activities right than it got wrong.

Each data point in Fig.5 represents a single simulation. The simulations are colored by the standard deviation of $noise$ in the calculation of $t_{sp}$. The set of simulations that are plotted were chosen as follows. If a set of parameters $[g_{\text{NMDA}}, g_{\text{GABA}_A}, N_{prf}, g_{\text{GABA}_B}]$ was found to give a successful simulation in any of the gamma modulation conditions (successful was defined as $AI > 0.7$), then the simulations with those parameters were plotted for all gamma modulation conditions. Of those plotted, only simulations with $AI > 0.7$ were included in the totals at the top of the graph, but all simulations plotted were included in the calculation of means, standard deviations, and regression of $AI$ against $GMI$. The error bars are standard error of the mean.

Simulations were written in C++. Numerical integration was performed using Euler's method and $\Delta t$ = 0.025 ms. Code is available upon request.